\documentclass{ws-procs9x6}

\usepackage{latexsym}
\usepackage{graphicx}
\usepackage{axodraw}

\newcommand{\be}{\begin{equation}}
\newcommand{\ee}{\end{equation}}
\newcommand{\ba}{\begin{eqnarray}}
\newcommand{\ea}{\end{eqnarray}}
\def\nn{\nonumber}

\newcommand{\Dslash}{{D \hskip -7pt /}}

\def\olmass{\begin{picture}(10,10)(10,10)
\PhotonArc(15,10)(10,0,180)2 5
\Line(0,10)(30,10)
\end{picture}}

\def\tlscbffig{\begin{picture}(1,30)(1,-5)
\Photon(-5,15)(35,15)2 5
\CArc(15,15)(20,0,180)
\CArc(15,15)(20,180,360)
\CArc(15,15)(22,0,180)
\CArc(15,15)(22,180,360)
\end{picture}}

\def\tlscbf{\begin{picture}(10,10)(10,10)
\Photon(4,15)(26,15)2 4
\CArc(15,15)(10,0,180)
\CArc(15,15)(10,180,360)
\CArc(15,15)(12,0,180)
\CArc(15,15)(12,180,360)
\end{picture}}

\def\tlspbf{\begin{picture}(10,10)(10,10)
\Photon(4,15)(26,15)2 4
\CArc(15,15)(10,0,180)
\CArc(15,15)(10,180,360)
\CArc(15,15)(11,0,180)
\CArc(15,15)(11,180,360)
\CArc(15,15)(12,0,180)
\CArc(15,15)(12,180,360)
\end{picture}}

\def\tlscfrph{\begin{picture}(10,10)(10,10)
\DashLine(5,15)(25,15){1}
\CArc(15,15)(10,0,180)
\CArc(15,15)(10,180,360)
\end{picture}}

\def\tlscfr{\begin{picture}(10,10)(10,10)
\Photon(5,15)(25,15)2 4
\CArc(15,15)(10,0,180)
\CArc(15,15)(10,180,360)
\end{picture}}

\def\olscbf{\begin{picture}(10,10)(10,10)
\CArc(15,15)(10,0,180)
\CArc(15,15)(10,180,360)
\CArc(15,15)(12,0,180)
\CArc(15,15)(12,180,360)
\end{picture}}

\def\olscfr{\begin{picture}(10,10)(10,10)
\CArc(15,15)(10,0,180)
\CArc(15,15)(10,180,360)
\end{picture}}

\def\olscbfv{\begin{picture}(10,10)(10,10)
\CArc(15,15)(10,0,180)
\CArc(15,15)(10,180,360)
\CArc(15,15)(12,0,180)
\CArc(15,15)(12,180,360)
\Vertex(15,4)2
\end{picture}}

\def\olscfrv{\begin{picture}(10,10)(10,10)
\CArc(15,15)(10,0,180)
\CArc(15,15)(10,180,360)
\Vertex(15,5)2
\end{picture}}

\begin{document}

\title{Self-duality, helicity and background field loopology}

\author{Gerald V. Dunne}

\address{Department of Physics\\
University of Connecticut\\
Storrs, CT 06269-3046, USA\\ 
E-mail: dunne@phys.uconn.edu}


\maketitle

\abstracts{I show that helicity plays an important role in the development of rules for computing higher loop effective Lagrangians. Specifically, the two-loop Heisenberg-Euler effective Lagrangian in quantum electrodynamics is remarkably simple when the background field has definite helicity (i.e., is self-dual). Furthermore, the two-loop answer can be derived essentially algebraically, and is naturally expressed in terms of one-loop quantities. This represents a generalization of the familiar ``integration-by-parts'' rules for manipulating diagrams involving free propagators to the more complicated case where the propagators are those for scalars or spinors in the presence of  a background field.}

%%%%%%%%%%%%%%%%%%%%%%%%%%%%%%%%%%%%%%%%%%%%%%%%%%%%%%%%%%%%%%%%%%%%%%%%
% You may put here the table of contents, just uncomment 3 lines below %
%%%%%%%%%%%%%%%%%%%%%%%%%%%%%%%%%%%%%%%%%%%%%%%%%%%%%%%%%%%%%%%%%%%%%%%%

%\vspace{0.5cm}
%\tableofcontents
%\newpage

\section{Basic Strategy}
\label{intro}

The basic strategy of the approach described in this talk is as follows:

$\bullet$  There has been dramatic progress in recent years in computing multiloop amplitudes in both gauge and gravitational theories, for small numbers of external legs\cite{glover,bern}. A fundamental role has been played by helicity amplitudes\cite{ttwu,mangano,bernhelicity}. Other key ideas\cite{glover,bern} include color decompositions, master diagrams, recurrence relations and differential equations, as well as the development of efficient algebraic manipulation programs such as FORM\cite{form}.

$\bullet$  Multiloop effective actions are generating functionals for multiloop amplitudes. They therefore encapsulate information about multiloop amplitudes with any number of external legs.
Unfortunately, very little is known about such  effective actions beyond one loop\cite{dunnekogan}. 

$\bullet$ The main goal here is to report on the development of computational rules for higher-loop effective Lagrangians, along the lines of the rules, such as ``integration-by-parts''\cite{chetyrkin}, developed for computing amplitude diagrams. This approach requires computing vacuum diagrams using propagators in background fields, rather than diagrams involving only free propagators.

$\bullet$ Here we show that for two-loop effective actions helicity plays an important role, leading to dramatic simplifications. In particular, the rules for manipulating diagrams involving propagators in a self-dual background are surprisingly simple\cite{gvdloops}. 

$\bullet$ In light of recent advances by Witten and collaborators\cite{witten} concerning a twistor-space approach to (tree and one-loop) helicity amplitudes, it would be interesting to see if these methods might be naturally extended to higher-loop amplitudes and possibly to higher-loop effective Lagrangians. As is familiar from instanton physics\cite{corrigan}, twistor space is the natural language  in which to describe propagators in self-dual backgrounds, so such an extension appears natural.

\section{Brief review of one loop results}
\label{1loop}

\begin{figure}[ht]
\centerline{\includegraphics[scale=0.5]{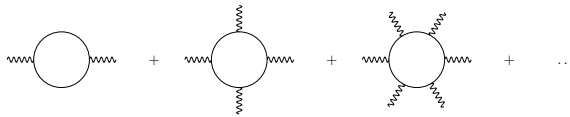}}
\caption{The diagrammatic perturbative expansion of the one loop effective action (\protect{\ref{action}}).}
\label{digexp}
\end{figure}

In classical field theory the Lagrangian encapsulates the relevant classical equations of motion and the symmetries of the system. In quantum field theory the effective Lagrangian encodes quantum corrections to the classical Lagrangian, corrections that are induced by quantum effects such as vacuum polarization. The seminal work of Heisenberg and Euler\cite{he}, and Weisskopf\cite{viki1} produced the paradigm for the entire field of effective Lagrangians by computing the nonperturbative, renormalized, one-loop effective action for quantum electrodynamics (QED) in a classical electromagnetic background of constant field strength:
\begin{equation}
S^{(1)}_{\rm spinor}=-i \ln \det (i\Dslash-m)=-\frac{i}{2}\ln \det(\Dslash^2+m^2)\, ,
\label{action}
\end{equation}
where the Dirac operator is $\Dslash =\gamma^\nu\left(\partial_\nu+ie A_\nu\right)$,  $A_\nu$ is a
fixed classical gauge potential with field strength tensor $F_{\mu\nu}=\partial_\mu A_\nu- \partial_\nu A_\mu$, and $m$ is the electron mass. This one-loop effective action has a natural perturbative expansion in powers of the external photon field $A_\mu$, as illustrated diagrammatically in Figure \ref{digexp}. By Furry's theorem (charge conjugation symmetry of QED), the expansion is in terms of even numbers of external photon lines.

In the low energy limit for the external photon lines, in which case the background field strength $F_{\mu\nu}$ could be taken to be constant, Heisenberg and Euler\cite{he} found a simple closed-form expression for the effective Lagrangian, which generates {\sl all} the perturbative diagrams in Figure \ref{digexp}:
\ba
{\mathcal L}_{\rm sp}^{(1)} =-\frac{1}{8\pi^2} 
\int_0^{\infty}\frac{d t}{t^3}
e^{-m^2 t}
\left\{
\frac{e^2a\,b\,t^2}{\tanh(e b t)\tan(e a t)} -1
-\frac{e^2 t^2}{3} (b^2-a^2)
\right\}.
\label{hesp}
\ea
Here $a$ and $b$ are related to the Lorentz invariants characterizing the background electromagnetic field strength:
$
a^2-b^2=\vec{E}^2-\vec{B}^2$, and $
a\, b= \vec{E}\cdot\vec{B}$. Weisskopf\cite{viki1} computed the analogous quantity for scalar QED
\begin{equation}
S^{(1)}_{\rm scalar}=\frac{i}{2}\ln \det(D_\mu^2+m^2)\, ,
\label{sqedaction}
\end{equation}
which involves the Klein-Gordon operator rather than the Dirac operator:
\ba
{\mathcal L}_{\rm sc}^{(1)} =\frac{1}{16\pi^2}  \int_0^{\infty}\frac{d t}{t^3}
e^{-m^2 t}
\left\{
\frac{e^2a\,b\,t^2}{ \sinh(e b t)\sin(e a t)} -1
+\frac{e^2 t^2}{6} (b^2-a^2)
\right\}.
\label{hesc}
\ea
The Heisenberg and Euler result (\ref{hesp}) leads immediately to a number of important physical insights and applications, such as the low energy limit of light-light scattering, and the existence of vacuum pair production in a background electric field.

\section{Two-Loop Heisenberg-Euler effective Lagrangian}
\label{twoloop}

\begin{figure}[ht]
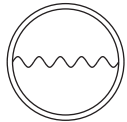

\centerline{\tlscbffig}
\caption{The two loop diagram for the two loop effective Lagrangian. The double line refers to a propagator in the presence of the constant background field, while the wavy line represents the internal virtual photon.}
\label{twoloopfig}
\end{figure}

In principle, the computation of the two-loop Heisenberg-Euler effective Lagrangian in QED is completely straightforward, as we only need to compute a single vacuum diagram (see Figure \ref{twoloopfig}) with an internal photon line and a single fermion (or scalar) loop, where these spinor (or scalar) propagators are in the presence of the background field. These background field propagators are well-known\cite{schwinger}. 
However, at two-loop we need to perform mass renormalization in addition to charge renormalization. Ritus\cite{ritus} found exact integral representations for the fully renormalized two loop Heisenberg-Euler effective Lagrangian in both spinor  and scalar QED. These are impressive computations, but unfortunately the answers are very complicated double-parameter integrals.

\subsection{Self-dual magic at two-loop}
\label{closedform}

Consider restricting the constant electromagnetic background to be self-dual: 
$
F_{\mu\nu}=\tilde{F}_{\mu\nu}
$, where $\tilde{F}_{\mu\nu}\equiv \frac{1}{2}\epsilon_{\mu\nu\rho\sigma}F^{\rho\sigma}$ is the standard dual electromagnetic field strength. 
Then the fully renormalized two-loop Heisenberg-Euler effective Lagrangian takes a remarkably simple closed-form (involving simple functions and no integrals!), for both spinor and scalar QED\cite{ds1}:
\begin{eqnarray}
{\mathcal L}_{\rm spinor}^{(2)}
&=&
-\alpha^2 \,\frac{f^2}{2\pi^2}\left[
3\,\xi^2 (\kappa)
-\xi'(\kappa)\right]\, ,
\label{2lsp}
\end{eqnarray}
\begin{eqnarray}
{\mathcal L}_{\rm scalar}^{(2)}
&=&
\alpha^2 \,\frac{f^2}{4\pi^2}\left[
\frac{3}{2}\,\xi^2 (\kappa)
-\xi'(\kappa)\right]\, .
\label{2lsc}
\end{eqnarray}
Here $f$ is the field strength parameter, $\frac{1}{4}F_{\mu\nu}F^{\mu\nu}=f^2$, and $\kappa$ is the natural  dimensionless parameter
$
\kappa\equiv \frac{m^2}{2e f} 
$.
The ubiquitous function $\xi(\kappa)$ is essentially the Euler digamma function $\psi(\kappa)=\frac{d}{d\kappa}\ln \Gamma(\kappa)$:
\begin{eqnarray}
\xi(\kappa)\equiv -\kappa\left(\psi(\kappa)-\ln(\kappa)+\frac{1}{2\kappa}\right) \, .
\label{xi}
\end{eqnarray}
The subtraction of the first two terms of the asymptotic expansion of $\psi(\kappa)$ correspond to renormalization subtractions, as shown below. It is also interesting to note that in such a self-dual background, the one-loop Heisenberg-Euler effective Lagrangians  (\ref{hesp}) and (\ref{hesc}) for spinor and scalar QED are also naturally expressed\cite{ds1} in terms of this same function $\xi(\kappa)$.

The dramatic simplicity of the two loop results (\ref{2lsp},\ref{2lsc}), compared to the complicated forms obtained by Ritus\cite{ritus},  raises three obvious  questions.
\begin{enumerate}
\item Why are these expressions so simple?

\item Why are the spinor and scalar expressions so similar?

\item Why is the particular function $\xi(\kappa)$ so special?
\end{enumerate}

\noindent The answers lie in the three-way relationship between self-duality, helicity and (quantum mechanical) supersymmetry.

\subsubsection{Simplicity of self-dual results}

Self-dual fields have definite helicity\cite{duffisham}. Indeed, the self-duality condition is just another way of writing the helicity projection:
\ba 
F_{\mu\nu}=\tilde{F}_{\mu\nu} \qquad \Leftrightarrow \qquad \sigma_{\mu\nu}F_{\mu\nu}\left(\frac{{\bf 1}+\gamma_5}{2}\right)=0\, .
\ea
For anti-self-dual fields the other helicity projection vanishes, so that the photon field has the opposite helicity. 

It is well-known that scattering amplitudes for external field lines with like helicities are particularly simple\cite{ttwu,mangano}. Since the effective action for a self-dual field is the generating function for like-helicity amplitudes, it is consistent that the effective action in a self-dual background should be simple. However, almost all of these helicity amplitude results are for massless particles on internal lines (but see for example\cite{bernmorgan}), while here we see a generalization to massive particles. 

A more prosaic reason for the simplicity of the two-loop expressions (\ref{2lsp}) and (\ref{2lsc}) is that for a self-dual field the square of the matrix $F_{\mu\nu}$ is proportional to the identity matrix:
$
F_{\mu\nu}F_{\nu\rho}=-f^2 \delta_{\mu\rho}
$.
This dramatically simplifies the propagators of spinors or scalars in the background field. For example, for a scalar particle
\begin{eqnarray}
G_{\rm scalar}(p)=\int_0^\infty \frac{dt}{\cosh^2(e f t)}\,e^{ -m^2 t
-\frac{p^2}{ef} \tanh(e f t)}\, .
\label{pmprop}
\end{eqnarray}
Note that this propagator is a function of  $p^2$, rather than of individual components of the momentum, which greatly simplifies the background field computations. The propagator satisfies a simple differential equation:
\ba
(p^2+m^2)G(p) = 1+\left(\frac{ef}{2}\right)^2\frac{\partial^2 G(p)}{\partial p_\mu^2}\, .
\label{mompropde}
\ea

\subsubsection{Similarity of spinor and scalar results in self-dual background}

Another consequence of the self-duality of the background is that the corresponding Dirac operator has a quantum mechanical supersymmetry. That is, apart from zero modes, the Dirac operator has the same spectrum (but with a multiplicity of 4) as the corresponding scalar Klein-Gordon operator\cite{thooft,jackiwrebbi}. At one-loop this implies \ba
{\mathcal L}^{(1)}_{\rm spinor}=-2{\mathcal L}^{(1)}_{\rm scalar}+\frac{1}{2} 
\left(\frac{ef}{2\pi}\right)^2\, \ln\left(\frac{m^2}{\mu^2}\right) \, ,
\label{1lsdconnection}
\ea
where $N_0=\left(\frac{ef}{2\pi}\right)^2$ is the zero mode number density. The logarithmic term in (\ref{1lsdconnection}) corresponds to the zero mode contribution. Renormalizing on-shell ({\it i.e.}, $\mu^2=m^2$), we find that the spinor and scalar effective Lagrangians (\ref{hesp}) and (\ref{hesc}) are proportional to one another for a self-dual background, in such a way that the SUSY combination vanishes: ${\mathcal L}^{(1)}_{\rm spinor}+2{\mathcal L}^{(1)}_{\rm scalar}=0$.

Now consider the implications of self-duality of the background at the two loop level. The two-loop the effective action is not simply a log determinant, so the situation is more complicated. Nevertheless, the quantum mechanical SUSY of the Dirac operator relates the spinor propagator to the scalar propagator via simple helicity projections, which has the consequence that after just doing the Dirac traces in the spinor two loop effective Lagrangian, one finds that  it can be written as the sum of two terms involving matrix elements of the scalar propagator. Moreover, these are the same two matrix elements of the scalar propagator that appear in the scalar QED effective Lagrangian, but with different numerical coefficients\cite{zmb}. This structure explains why the two loop answers (\ref{2lsp}) and (\ref{2lsc}) for spinor and scalar QED have such a similar form, involving just two terms with different numerical coefficients.

\subsubsection{Significance of $\xi$ and $\xi^\prime$}

An important question to address is why are the simple expressions (\ref{2lsp}) and (\ref{2lsc}) for the two loop Heisenberg-Euler effective Lagrangians in a self-dual background expressed in terms of the particular function $\xi(\kappa)$ and its derivative [recall that $\xi$ was defined in (\ref{xi}) as essentially the Euler digamma function]. The first hint comes from the following facts that for a self-dual background the following scalar propagator loops, evaluated using the background field propagator (\ref{pmprop}), are simply related to the $\xi(\kappa)$ function:
\ba
\mbox{\olscbf}\quad -\quad\mbox{\olscfr} \quad&\equiv &\int \frac{d^4p}{(2\pi)^4}\left[G(p)-G_0(p)\right]=-\frac{m^2}{(4\pi)^2} \,\frac{ \xi(\kappa)}{\kappa}\ ,
\label{xiint}\\[2mm]
\mbox{\olscbfv}\quad -\quad\mbox{\olscfrv} \quad &\equiv &\int \frac{d^4p}{(2\pi)^4}\left[(G(p))^2-(G_0(p))^2\right]= \frac{\xi^\prime(\kappa)}{(4\pi)^2}\ .
\label{xiprint}
\ea
Here the double lines refer to scalar propagators in the self-dual background and the single line is the free scalar propagator, while the dot on a propagator refers to the propagator being squared. Thus, $\xi$ and $\xi^\prime$ are natural one loop traces for the propagator in a self-dual background.

Given (\ref{xiint}) and (\ref{xiprint}), we can write the closed-form expressions (\ref{2lsp}) and (\ref{2lsc})  for the two loop effective Lagrangians in diagrammatic form:
\ba
{\rm spinor\,\,QED} &:&\nn\\
\quad\Bigg[~~\mbox{\tlspbf}\quad - \hskip.3cm \mbox{\tlscfr}~~\Bigg]&=&-6\,e^2\, 
\Bigg[~~\mbox{\olscbf}\quad - \hskip.3cm \mbox{\olscfr} ~~\Bigg]^2 +\frac{(ef)^2}{2\pi^2} \Bigg[~~\mbox{\olscbfv}\quad - \hskip.3cm  \mbox{\olscfrv}~~\Bigg]\,,\nn\\
\label{2lspdig}\\
{\rm scalar\,\,QED}&:&\nn\\
 \quad\Bigg[~~\mbox{\tlscbf}\quad - \quad \mbox{\tlscfr}~~\Bigg]&=&\frac{3}{2}\,e^2\,
 \Bigg[~~\mbox{\olscbf}\quad -  \hskip.3cm \mbox{\olscfr}~~\Bigg]^2 -\frac{(ef)^2}{4\pi^2}  \Bigg[~~\mbox{\olscbfv}\quad - \hskip.3cm  \mbox{\olscfrv}~~\Bigg]\,.\nn\\
\label{2lscdig}
\ea
Here the notation is that the triple line loop on the LHS of (\ref{2lspdig}) refers to a spinor propagator in a self-dual background, while the double-line loops [including those on the RHS of  (\ref{2lspdig})] refer to a scalar propagator in the self-dual background. This shows the remarkable result that the two loop fully renormalized answers are expressed naturally in terms of one loop quantities. Qualitatively, we can write:
\ba
%\fbox{\begin{minipage}{6.5cm}{$$
{\rm two\,\,loop}=\left({\rm one\,\,loop}\right)^2+\left({\rm one\,\,loop}\right)
%$$}
%\end{minipage}}
\label{2l1l}
\ea
Interestingly, such a relation with two loop quantities being expressed as squares of one loop quantities plus a one loop remainder has been found recently\cite{abdk} in the amplitudes of 4 dimensional super Yang-Mills theory, which is a very different theory from QED. Also, the same function $\xi(\kappa)$, and its derivatives appear naturally in recent studies of ${\mathcal N}=2$ SUSY QED and YM effective Lagrangians at two-loop\cite{kuzenko}. This suggests something deeper is at work here.

\section{Background field loopology}
\label{intbyparts}

It is natural to ask if the remarkable simplifications of the two loop results for a self-dual background might extend to even higher loops. To go beyond two loops one should take advantage of the great progress that has been made recently in understanding the structure of higher-loop quantum field theory (without background fields). The general strategy is to manipulate diagrams to reduce the number to a much smaller set of so-called "master diagrams" which need to be computed. This has led, for example, to many new two-loop results for QCD scattering amplitudes\cite{glover,bern}. 
I conclude this talk with some comments and speculations about how these techniques can be extended to incorporate background fields\cite{gvdloops}.

Indeed, we can go further than the qualitative statement (\ref{2l1l}) and derive the results (\ref{2lspdig}) and (\ref{2lscdig}) by algebraic means. First, we identify the source of the coefficient factors $-6e^2$ and $\frac{3}{2}e^2$ which appear in front of the $\left({\rm one\,\,loop}\right)^2$ terms in (\ref{2lspdig}) and (\ref{2lscdig}). Note that 
in free QED ({\it i.e.},  with no background field) it is a straightforward exercise to show that in 4 dimensions
\ba
{\rm spinor\,\, QED} : \qquad \mbox{\tlscfr} \quad &=& -6\,e^2\,\bigg[~~\mbox{\olscfr}~~\bigg]^2 \,,
\label{sp2}\\ \nn\\
{\rm scalar\,\, QED} : \qquad\mbox{\tlscfr} \quad &=& ~\frac{3}{2}\,e^2\,\bigg[~~\mbox{\olscfr}~~\bigg]^2\,.
\label{sc2}
\ea
(The loop on the RHS is a scalar loop in each case.)  Thus, the two loop free vacuum diagram can be expressed in terms of a simpler one loop diagram. These results can either be derived by computing each side using dimensional regularization, or a quicker proof follows from an integration-by-parts argument (see below). Notice that the coefficients of the $\left({\rm one\,\,loop}\right)^2$ parts in this free case are exactly the same as the corresponding coefficients in the background field expressions (\ref{2lspdig}) and (\ref{2lscdig}), for both spinor and scalar QED. This is no accident, as I now illustrate for the case of scalar QED (for spinor QED the argument is similar).

Consider the derivation of (\ref{sc2}) using dimensional regularization and integration-by-parts \cite{chetyrkin}. By purely algebraic manipulations
\ba 
\tlscfr \hskip.3cm &=&\frac{e^2}{2} \int \frac{d^dp\, d^dq}{(2\pi)^{2d}}\, \frac{(p+q)^2}{(p-q)^2(p^2+m^2)(q^2+m^2)}\nn \\ \nn\\
&=& \frac{e^2}{2} \int \frac{d^dp\, d^dq}{(2\pi)^{2d}}\, \frac{\left[-(p-q)^2+2(p^2+m^2)+2(q^2+m^2)-4 m^2\right]}{(p-q)^2(p^2+m^2)(q^2+m^2)}\nn \\ \nn \\
&=& -\frac{e^2}{2}\bigg[~~\olscfr~~\bigg]^2 -2 e^2 m^2  
\bigg[~~\tlscfrph~~\bigg]\ ,
\label{freemanip}
\ea
where the dotted line denotes a massless scalar propagator. The first term has been written as the square of a one loop diagram but the second term is apparently still two loop. However, using simple integration-by-parts manipulations, this two loop diagram can also be written as a square of a one loop diagram:
\ba 
0 &=& \int \frac{d^dp\, d^dq}{(2\pi)^{2d}}\,\frac{\partial}{\partial p_\mu}  \left[\frac{(p-q)_\mu}{(p-q)^2(p^2+m^2)(q^2+m^2)}\right]\nn \\ \nn\\
&=& (d-2) \bigg[~~\tlscfrph~~\bigg] - \int \frac{d^dp\, d^dq}{(2\pi)^{2d}}\, \frac{2p\cdot (p-q)}{(p-q)^2(p^2+m^2)^2(q^2+m^2)}\nn \\ \nn \\
&=& (d-2) \left[\hskip .3cm \tlscfrph \hskip .3cm \right] - \int \frac{d^dp\, d^dq}{(2\pi)^{2d}}\, \frac{[(p-q)^2+(p^2+m^2)-(q^2+m^2)]}{(p-q)^2(p^2+m^2)^2(q^2+m^2)}\nn \\ \nn \\
&=& (d-3) \bigg[~~\tlscfrph~~\bigg] -  \bigg[~~\olscfrv~~\bigg]  \bigg[~~\olscfr~~\bigg]
\nn \\ \nn \\
&=&(d-3) \bigg[~~\tlscfrph~~\bigg] -  \frac{(d-2)}{2m^2}\bigg[~~\olscfr~~\bigg]^2\,. \label{bp}
\ea
Thus, the apparently $2$-loop term on the RHS of (\ref{freemanip}) is a square of a one loop diagram, leading to
\ba 
\tlscfr \hskip .3cm = \frac{e^2}{2}\left(\frac{d-1}{d-3}\right)\bigg[~~\olscfr~~\bigg]^2\,.
\label{free1lsq}
\ea
This reduces to (\ref{sc2}) in $d=4$, and I stress that this result has been derived without doing any integrations, only making simple algebraic manipulations on the integrands.

Now consider the analogous manipulation in a self-dual background\cite{gvdloops}. First, we extend the scalar propagator in a self-dual background to arbitrary dimensions by taking multiple copies of the block diagonal structure of $F_{\mu\nu}$.  This is equivalent to dimensional regularization in the worldline formalism\cite{csreview}. Then the scalar propagator (\ref{pmprop}) becomes
\begin{eqnarray}
G(p)=\int_0^\infty \frac{dt}{\cosh^{d/2}(eft)} \, e^{-m^2 t-\frac{p^2}{ef}\tanh(eft)}\ .
\label{dsdscprop}
\end{eqnarray}
Then we can repeat the algebraic steps in (\ref{freemanip}) to obtain
\ba 
\tlscbf \hskip.3cm &=& \frac{e^2}{2} \int \frac{d^dp\, d^dq}{(2\pi)^{2d}}\, \frac{1}{(p-q)^2}\left\{ (p+q)^2G(p)G(q)-(ef)^2 \frac{\partial G(p)}{\partial p_\mu}  \frac{\partial G(q)}{\partial q_\mu} \right\}\nn  \\
\nn\\
&=& \frac{e^2}{2}\left(\frac{d-1}{d-3}\right)\bigg[~~\olscbf~~~\bigg]^2+O(f^2) \ ,
\label{bfmanip}
\ea
where we have chosen to isolate this particular coefficient of the square of the one loop propagator trace motivated by the free-field result (\ref{free1lsq}). 

The advantage of the manipulation in (\ref{bfmanip}) is that it makes the mass renormalization (which was a very difficult part of previous two loop computations) almost trivial. To see this we subtract the free field two loop diagram from the background field diagram
\ba
 \bigg[~~~\mbox{\tlscbf}\hskip .3cm- \hskip .3cm\mbox{\tlscfr}~~\bigg]=\frac{e^2}{2}\left( \frac{d-1}{d-3}\right) \left\{\bigg[~~~\olscbf~~~\bigg]^2- \bigg[~~\olscfr~~\bigg]^2 \right\}+O(f^2)\ ,
 \nn\\
\label{diff}
\ea
and then simply complete the square in the first terms:
\ba 
\hskip -0.4cm \bigg[~~\olscbf~~\bigg]^2\!\!\!- \bigg[~~\olscfr~~\bigg]^2\!\!\!=
\bigg[~~\olscbf~~ - ~~\olscfr~~\bigg]^2\!\!\!+2\bigg[~~\olscfr~~\bigg]
\bigg[~~\olscbf~~ - ~~\olscfr~~\bigg]\,.
\label{complete}
\ea
The cross-term in (\ref{complete}) is immediately identified with the mass renormalization because 
\ba
\delta m^2=\bigg[\hskip .4cm \olmass \hskip .4cm\bigg]_{p^2=-m^2}=e^2\left( \frac{d-1}{d-3}\right) \bigg[\hskip .3cm\olscfr\hskip .3cm \bigg] \ ,
\label{massshift}
\ea
which can also be derived algebraically. Furthermore, 
\ba 
 \bigg[ \hskip .3cm \olscbf\hskip .3cm -\hskip .3cm \olscfr\hskip .2cm \bigg] = -\frac{\partial {\mathcal L}^{(1)}}{\partial (m^2)}+(f^2\,\,{\rm term})\ .
%-\frac{e^2 f^2}{m^2 (4\pi)^2}\left(\frac{m^2}{4\pi}\right)^{\frac{d-4}{2}}\frac{d}{12} \Gamma\left(3-%\frac{d}{2}\right)
\label{der1l}
\ea
The $f^2$ term in (\ref{der1l}) contributes to the charge renormalization, and so (\ref{diff}) can be written as
\ba
\bigg[\hskip .3cm \mbox{\tlscbf}\hskip .3cm- \hskip .3cm\mbox{\tlscfr}\hskip .3cm\bigg]&=&\frac{e^2}{2}\left( \frac{d-1}{d-3}\right) \bigg[\hskip .3cm \mbox{\olscbf}\hskip .3cm- \hskip .3cm \mbox{\olscfr}\hskip .3cm \bigg]^2\nn\\
&&- \delta m^2\, \frac{\partial {\mathcal L}^{(1)}}{\partial (m^2)}
+O(f^2)\ .
\label{massrenorm}
\ea
Observe that the first term is now completely finite, so that we can set $d=4$, and by (\ref{xiint}) we obtain the first term of the final answer (\ref{2lscdig}), the $({\rm one\,\,loop})^2$ piece, without doing any integrals at all! The second term is manifestly the mass renormalization term, and so is absorbed by mass renormalization. The only remaining divergence can be proportional to the bare Maxwell Lagrangian $f^2$, which is then subtracted by charge renormalization. It is simple to isolate and subtract this piece, leaving an $O(f^4)$ term, whose kernel in the $d\to 4$ limit reduces to a momentum delta function:
\ba
\hskip -1cm O(f^4)&{\buildrel {d\to 4} \over \longrightarrow}&-4\pi^2 e^2 (ef)^2 \int \frac{d^4pd^4q}{(2\pi)^8}\left[G(p) G(q)-G_0(p) G_0(q)\right] \, \delta(p-q)\nonumber\\ \nonumber\\
&=&
-\frac{e^2}{4\pi^2} \, (ef)^2\bigg[ \hskip .4cm \mbox{\olscbfv}\hskip .3cm -\hskip .3cm\mbox{\olscfrv}\hskip .3cm\bigg]\ ,
\label{xippiece}
\ea
where in the last step we have used (\ref{xiprint}). Thus, we have derived the diagrammatic form (\ref{2lscdig}) of the fully renormalized two loop scalar QED effective Lagrangian in a self-dual background by essentially algebraic manipulations. It would be interesting to develop these background field ``integration-by-parts" rules into a fully systematic set of rules that might be applied to even higher loop order.

\section{Conclusions}
\label{conclusions}

To conclude, I reiterate that self-duality appears to be playing a remarkable simplifying role in the computation of higher-loop effective Lagrangians. Analogously, self-duality is an important simplifying principle in the computation of higher-loop amplitudes. This talk is a first attempt to bring together these two aspects of higher-loop computations. The main goal is the development of computational techniques for computing higher-loop vacuum diagrams involving propagators in the presence of background fields, generalizing the integration-by-parts rules developed for higher-loop diagrams involving free propagators. Such background field vacuum diagrams are the building blocks of higher-loop effective Lagrangians.  For a self-dual background this can be done explicitly to the two-loop level, and it is suggested that expansion about the self-dual background will provide a starting point for more general backgrounds. Furthermore, for a self-dual background, such algebraic rules for manipulating diagrams should facilitate the computation of even higher-loop effective Lagrangians. Finally, it would be interesting to investigate whether Kreimer's Hopf algebra structure underlying Feynman diagrams\cite{dirk} can be usefully applied to these background field computations.

\section*{Acknowledgments}
I thank Christian Schubert and Holger Gies for collaboration, and  the US DOE for support through grant DE-FG02-92ER40716.


\begin{thebibliography}{0}

\bibitem{glover}
E.~W.~Glover,
%``Progress in NNLO calculations for scattering processes,''
Nucl.\ Phys.\ Proc.\ Suppl.\  {\bf 116}, 3 (2003)
[arXiv:hep-ph/0211412].
%%CITATION = HEP-PH 0211412;%%

\bibitem{bern}
Z.~Bern,
%``Recent progress in perturbative quantum field theory,''
Nucl.\ Phys.\ Proc.\ Suppl.\  {\bf 117}, 260 (2003)
[arXiv:hep-ph/0212406].
%%CITATION = HEP-PH 0212406;%%

\bibitem{ttwu} 
R. Gastmans and T. T. Wu, 
{\it The ubiquitous photon: helicity method for QED and QCD}, (Oxford Univerity Press,
New York, 1990).

\bibitem{mangano} 
M.~L.~Mangano and S.~J.~Parke,
%``Multiparton Amplitudes In Gauge Theories,''
Phys.\ Rept.\  {\bf 200}, 301 (1991);
%%CITATION = PRPLC,200,301;%%

\bibitem{bernhelicity}
Z.~Bern, G.~Chalmers, L.~J.~Dixon and D.~A.~Kosower,
%``One loop N gluon amplitudes with maximal helicity violation via collinear
%limits,''
Phys.\ Rev.\ Lett.\  {\bf 72}, 2134 (1994)
[arXiv:hep-ph/9312333].
%%%CITATION = HEP-PH 9312333;%%


\bibitem{form}
J.~A.~M.~Vermaseren,
``New features of FORM,''
arXiv:math-ph/0010025.
%%CITATION = MATH-PH 0010025;%%


\bibitem{dunnekogan}
G.~V.~Dunne, ``Heisenberg-Euler Effective Lagrangians: Basics and Extensions'', arXiv:hep-th/0406216, to appear in Ian Kogan Memorial Collection, {\it From Fields to Strings: Circumnavigating Theoretical Physics'}, M. Shifman, A. Vainshtein and J. Wheater (Eds), (World Scientific, Singapore).



\bibitem{chetyrkin}
K.~G.~Chetyrkin and F.~V.~Tkachov,
%``Integration By Parts: The Algorithm To Calculate Beta Functions In 4 Loops,''
Nucl.\ Phys.\ B {\bf 192}, 159 (1981).
%%CITATION = NUPHA,B192,159;%%

\bibitem{gvdloops}
G.~V.~Dunne,
%``Two-loop diagrammatics in a self-dual background,''
JHEP {\bf 0402}, 013 (2004)
[arXiv:hep-th/0311167].
%%CITATION = HEP-TH 0311167;%%


\bibitem{witten}
E.~Witten,
``Perturbative gauge theory as a string theory in twistor space,''
arXiv:hep-th/0312171;
%%CITATION = HEP-TH 0312171;%%
F.~Cachazo, P.~Svrcek and E.~Witten,
``MHV vertices and tree amplitudes in gauge theory,''
arXiv:hep-th/0403047;
%%CITATION = HEP-TH 0403047;%%
``Twistor space structure of one-loop amplitudes in gauge theory,''
arXiv:hep-th/0406177.
%%CITATION = HEP-TH 0406177;%%


\bibitem{corrigan}
E.~Corrigan, D.~B.~Fairlie, S.~Templeton and P.~Goddard,
%``A Green's Function For The General Selfdual Gauge Field,''
Nucl.\ Phys.\ B {\bf 140}, 31 (1978);
%%CITATION = NUPHA,B140,31;%%
N.~H.~Christ, E.~J.~Weinberg and N.~K.~Stanton,
%``General Self-Dual Yang-Mills Solutions,''
Phys.\ Rev.\ D {\bf 18}, 2013 (1978).
%%CITATION = PHRVA,D18,2013;%%



\bibitem{he}
W. Heisenberg and H. Euler,
% ``Consequences of Dirac's Theory of Positrons'', 
 Z. Phys. {\bf 98} (1936) 714.

\bibitem{viki1} V. Weisskopf, 
% ``The electrodynamics of the vacuum based on the quantum theory of the electron'', 
 Kong. Dans. Vid. Selsk. Math-fys. Medd. XIV No. 6 (1936)
% ; English translation in: {\it Early Quantum Electrodynamics: A Source Book}, A. I. Miller, (Cambridge University Press, 1994).


\bibitem{schwinger}
J. Schwinger, 
%``On gauge invariance and vacuum polarization'', 
Phys. Rev. {\bf 82} (1951) 664.



\bibitem{ritus}
V. I. Ritus, 
%``Lagrangian of an intense electromagnetic
%field and quantum electrodynamics at short distances'', 
%Zh. Eksp. Teor. Fiz {\bf 69} (1975) 1517 [
Sov. Phys. JETP {\bf 42} (1975) 774;
%].
%
%\bibitem{ritusscal}
%V. I. Ritus, 
%``Connection between strong-field quantum electrodynamics
%with short-distance quantum electrodynamics'',
%Zh. Eksp. Teor. Fiz {\bf 73} (1977) 807
%[
Sov. Phys. JETP {\bf 46} (1977) 423.

\bibitem{ds1}
G.~V.~Dunne and C.~Schubert,
%``Closed-form two-loop Euler-Heisenberg Lagrangian in a self-dual 
%background,'' 
Phys.\ Lett.\ B {\bf 526}, 55 (2002)
[arXiv:hep-th/0111134];
%%%CITATION = HEP-TH 0111134;%%
%\bibitem{ds2}
%G.~V.~Dunne and C.~Schubert,
%``Two-loop self-dual Euler-Heisenberg Lagrangians. I: Real part and 
%helicity amplitudes,'' 
JHEP {\bf 0208}, 053 (2002)
[arXiv:hep-th/0205004];
%%CITATION = HEP-TH 0205004;%%
%
%\bibitem{ds3}
%G.~V.~Dunne and C.~Schubert,
%%``Two-loop self-dual Euler-Heisenberg Lagrangians. II: Imaginary part 
%%and Borel analysis,''
 JHEP {\bf 0206}, 042 (2002)
[arXiv:hep-th/0205005].
%%CITATION = HEP-TH 0205005;%%


\bibitem{duffisham}
M.~J.~Duff and C.~J.~Isham,
%``Selfduality, Helicity, And Supersymmetry: The Scattering Of Light By Light,''
Phys.\ Lett.\ B {\bf 86}, 157 (1979);
%%CITATION = PHLTA,B86,157;%%
%``Selfduality, Helicity, And Coherent States In Nonabelian Gauge Theories,''
Nucl.\ Phys.\ B {\bf 162}, 271 (1980).
%%CITATION = NUPHA,B162,271;%%

\bibitem{bernmorgan}
Z.~Bern and A.~G.~Morgan,
%``Massive Loop Amplitudes from Unitarity,''
Nucl.\ Phys.\ B {\bf 467}, 479 (1996)
[arXiv:hep-ph/9511336].
%%CITATION = HEP-PH 9511336;%%



\bibitem{thooft}
G.~'t Hooft,
%``Computation Of The Quantum Effects Due To A Four-Dimensional 
%Pseudoparticle,'' 
Phys.\ Rev.\ D {\bf 14}, 3432 (1976)
[Erratum-ibid.\ D {\bf 18}, 2199 (1978)].
%%CITATION = PHRVA,D14,3432;%%

\bibitem{jackiwrebbi}
R.~Jackiw and C.~Rebbi,
%``Degrees Of Freedom In Pseudoparticle Systems,''
%Phys.\ Lett.\ B {\bf 67}, 189 (1977);
%%%CITATION = PHLTA,B67,189;%%
%``Spinor Analysis Of Yang-Mills Theory,''
Phys.\ Rev.\ D {\bf 16}, 1052 (1977).
%%CITATION = PHRVA,D16,1052;%%

%\bibitem{dadda}
%A.~D'Adda and P.~Di Vecchia,
%%``Supersymmetry And Instantons,''
%Phys.\ Lett.\ B {\bf 73}, 162 (1978).
%%%CITATION = PHLTA,B73,162;%%

\bibitem{zmb}
G.~V.~Dunne, H.~Gies and C.~Schubert,
%``Zero modes, beta functions and IR/UV interplay in higher-loop QED,''
JHEP {\bf 0211}, 032 (2002)
[arXiv:hep-th/0210240].
%%CITATION = HEP-TH 0210240;%%

\bibitem{abdk}
C.~Anastasiou, Z.~Bern, L.~Dixon and D.~A.~Kosower,
%``Planar amplitudes in maximally supersymmetric Yang-Mills theory,''
Phys.\ Rev.\ Lett.\  {\bf 91}, 251602 (2003)
[arXiv:hep-th/0309040].
%%CITATION = HEP-TH 0309040;%%




\bibitem{kuzenko}
S.~M.~Kuzenko and I.~N.~McArthur,
%``Low-energy dynamics in N = 2 super QED: Two-loop approximation,''
%JHEP {\bf 0310}, 029 (2003)
%[arXiv:hep-th/0308136];
%%%CITATION = HEP-TH 0308136;%%
Phys.\ Lett.\ B {\bf 591}, 304 (2004)
[arXiv:hep-th/0403082],
%%CITATION = HEP-TH 0403082;%%
``Relaxed super self-duality and N = 4 SYM at two loops,''
arXiv:hep-th/0403240.
%%CITATION = HEP-TH 0403240;%%


\bibitem{csreview}
C. Schubert, 
%``Perturbative quantum field theory in the string-inspired formalism'', 
Phys. Rept. {\bf 355} (2001) 73, hep-th/0101036.







\bibitem{dirk}
D.~Kreimer,
%``On the Hopf algebra structure of perturbative quantum field theories,''
Adv.\ Theor.\ Math.\ Phys.\  {\bf 2}, 303 (1998)
[arXiv:q-alg/9707029];
%%CITATION = Q-ALG 9707029;%%
A.~Connes and D.~Kreimer,
%``Renormalization in quantum field theory and the Riemann-Hilbert  problem. I: The Hopf algebra structure of graphs and the main theorem,''
Commun.\ Math.\ Phys.\  {\bf 210}, 249 (2000)
[arXiv:hep-th/9912092];
%%CITATION = HEP-TH 9912092;%%
%D. Kreimer, {\it Knots and Feynman Diagrams}, (Cambridge Univ Press, 2000).


\end{thebibliography}
\end{document}